\documentclass[onecolumn,showpacs,preprintnumbers,amsmath,amssymb]{revtex4}
\usepackage{amssymb,amsmath,amsthm,mathtools}
\usepackage{hyperref}
\usepackage{url,graphicx,epstopdf} 
\usepackage{graphicx}
\usepackage{dcolumn}
\usepackage{bm}
\usepackage{epsfig}
\usepackage{subfigure}
\usepackage{xcolor}

\newcommand{\bi}{\begin{itemize}}
\newcommand{\ei}{\end{itemize}}
\newcommand{\be}{\begin{eqnarray}}
\newcommand{\ee}{\end{eqnarray}}
\newcommand{\beq}{\begin{equation}}
\newcommand{\eeq}{\end{equation}}
\newcommand{\bbmatrix}{\left( \begin{array}}
\newcommand{\eematrix}{\end{array} \right)}

\begin{document} 

\title{Application of Pontryagin's Maximum Principle to Quantum Metrology in Dissipative Systems } 

\author{Chungwei Lin$^1$\footnote{clin@merl.com}, Yanting Ma$^1$} 

\affiliation{$^1$Mitsubishi Electric Research Laboratories, 201 Broadway, Cambridge, MA 02139, USA 
} 

\author{Dries Sels$^{2,3}$}

\affiliation{
$^2$Department of physics, New York University, New York City, NY 10003, USA \\
$^3$Center for Computational Quantum Physics, Flatiron Institute, 162 5th Ave, New York, NY 10010, USA\footnote{The Flatiron Institute is a division of the Simons Foundation.}
} 

\begin{abstract}
Optimal control theory, also known as Pontryagin's Maximum Principle, is applied to the quantum parameter estimation in the presence of decoherence. An efficient procedure is devised to compute the gradient of quantum Fisher information with respect to the control parameters and is used to construct the optimal control protocol. The proposed procedure keeps the control problem in the time-invariant form so that both first-order and second-order optimality conditions derived from Pontryagin's Maximum Principle apply; the second-order condition turns out to be crucial  when the optimal control contains singular arcs. Concretely we look for the optimal control that maximizes quantum Fisher information for ``twist and turn'' problem. We find that the optimal control is singular without dissipation but can become unbounded  once the quantum decoherence is introduced. An amplitude constraint is needed to guarantee a bounded solution. With quantum decoherence, the maximum quantum Fisher information happens at a finite time due to the decoherence, and the asymptotic value depends on the specific decoherence channel and the control of consideration. 
\end{abstract}

\maketitle 

\section{Introduction}  

Open-loop optimal control theory, also known as Pontryagin's Maximum Principle (PMP) \cite{book:Pontryagin, Sussmann_87_01, book:Luenberger, book:GeometricOptimalControl}, is a powerful tool in classical control theory. It deals with minimizing a terminal cost function conditioned on the dynamics that contains control variables parametrized as a function of time. The core of PMP is the calculus of variations. In particular, it provides an computationally efficient procedure to evaluate the gradient of the terminal cost with respect to the control variables, as well as optimality conditions that constrain the general behavior of an optimal control. 
Many problems in modern quantum technology \cite{Neilson_book, book:Kaye_book, Shor:1997:PAP:264393.264406,  RevModPhys.66.481, RevModPhys.77.513, LIGO_2011} naturally fit this framework. 
Important examples include quantum state preparation \cite{PhysRevA.97.062343,  ahmed19, PhysRevA.74.022306, PhysRevX.8.021012, PhysRevLett.112.047601, PhysRevA.90.013409, PhysRevA.93.013851} where the terminal cost is the overlap to the known target state, the ``continuous-time'' variation-principle based quantum computation \cite{Farhi_00, PhysRevLett.103.080502, PhysRevA.90.052317, Peruzzo-2014, PhysRevX.6.031007} where the terminal cost is the ground-state energy, and quantum parameter estimation (quantum metrology) \cite{book:Helstrom, book:Holevo, PhysRevLett.96.010401, Brif_2010, Glaser-2015, AdvancesQuantumMetrology_2011, PhysRevX.6.031033, PhysRevA.96.040304,  RevModPhys.90.035005, PhysRevX.10.031003, PhysRevLett.110.053002, Sekatski2017quantummetrology, PhysRevA.96.032310, doi:10.1116/5.0006785, PhysRevLett.125.020402} where the cost function is the Fisher information.  In a more general context, optimal control has applied to the stabilization of ultracold molecules \cite{PhysRevA.70.013402} and cooling of quantum systems \cite{PhysRevA.82.063422, doi:10.1137/100818431, PhysRevA.87.043607}. 

Maximal quantum Fisher information (QFI) has been used for optimal estimation of Hamiltonian parameters  \cite{Pang-2017,  PhysRevLett.123.260505, PhysRevLett.124.060402, Koczor_2020}. Numerically, the Fisher information can be optimized by e.g. GRAPE (GRadient Ascent Pulse Engineering \cite{Khaneja-2005}) both for single and multiple-parameter estimations in the presence of noise \cite{PhysRevA.96.012117, PhysRevA.96.042114, PhysRevResearch.2.033396, Liu_2019}. Recently we provide an alternate procedure \cite{PhysRevA.103.052607} to compute the gradient of quantum and classical Fisher information for closed quantum systems. 
In this work, we extend our previous work to take the quantum decoherence into account. 
To consider the decoherence effect requires density matrix (DM) as dynamical variables, and the evaluation of QFI becomes more involved compared with that based on wave functions. As a first step we develop an efficient procedure, based on the augmented dynamics \cite{PhysRevA.103.052607}, that allows us to stably compute the quantities introduced by PMP. With this numerical tool, we investigate the decoherence effect by solving the ``twist and turn'' problem \cite{RevModPhys.90.035005, Hayes_2018, PhysRevLett.124.060402} under various decoherence channels. 
In the presence of quantum decoherence, there are competing mechanisms for QFI. On the one hand, QFI increases
upon elongated interrogation 
which favors a long evolution time. On the other hand, QFI decreases upon reduced coherence which favors a short evolution time. One naturally expects a finite optimal evolution time that gives the maximum QFI -- this is roughly true but a more precise answer hinges on the interplay between the decoherence-free subspace \cite{PhysRevA.101.022320} and the control of consideration. In terms of control, we find that the optimal control is singular without decoherence, and an amplitude constraint turns out to be necessary for a bounded solution in the presence of decoherence. 

The paper is organized as follows. In Section II we present  the problem and provide the analysis framework based on PMP. First-order and second-order necessary conditions are explicitly stated. The procedure that efficiently computes the switching function is described in detail. In Section III, we give the optimal controls and the corresponding QFI for the two-spin problems under different dissipation channels. The numerical solutions are checked against the first and second order optimality conditions.
Conclusions are given in Section IV. Appendixes are provided to fill some intermediate steps skipped in the main text.

\section{Problem statement and Optimal control}

\subsection{Twist and turn Hamiltonian and Quantum Fisher information} 
The master equation of the density matrix $\hat{\rho}$ for the most general type of Markovian and time-homogeneous systems can be expressed as 
\begin{equation}
\frac{\partial}{\partial t} \hat{\rho} 
= -i [\hat{H}(t), \hat{\rho}] + \left\{ \frac{d \hat{\rho} }{dt} \right\}_\text{channel},
\end{equation}
which contains a unitary and a dissipative part. The unitary dynamics considered here is  the ``twist and turn'' Hamiltonian \cite{RevModPhys.90.035005, Hayes_2018, PhysRevLett.124.060402}
\begin{equation}
\hat{H} (t) = \chi \hat{J}_z^2 + \omega \hat{J}_z + u(t) \hat{J}_x,
\label{eqn:tt_Hamil}
\end{equation} 
with $[\hat{J}_i, \hat{J}_j] = i\, \epsilon_{ijk} \hat{J}_k$ ($i=x,y,z$). The initial state is chosen to be the non-entangled maximum-eigenvalue state of $\hat{J}_x$, also known as the spin coherent state, and will be denoted as $| \Psi_\text{coh-x} \rangle$. The relevant physical realizations include  interacting (generalized) spins \cite{PhysRevLett.123.260505, PhysRevX.10.031003}, the two-arm interferometer \cite{PhysRevA.85.022322, PhysRevA.33.4033}, and superradiance \cite{PhysRev.93.99, GROSS1982301}. In this work Eq.~\eqref{eqn:tt_Hamil} will be represented a set of $N$ all-to-all interacting spins where $\hat{J}_i = \sum_{n=1}^N \frac{\sigma_i}{2}$ ($i=x,y,z$ and $\sigma$'s are Pauli matrices). 
The dissipative part $ \left\{ \frac{d \hat{\rho} }{dt} \right\}_\text{channel}$ is classified by the ``decoherence channel'' whose explicit form will be specified shortly. 
In Eq.~\eqref{eqn:tt_Hamil}, $\hat{J}_z^2$ is the source of entanglement and referred to as the ``twist'' term; $\hat{J}_x$ the external control and the ``turn'' term; $\omega$ the parameter to estimate. The optimal control problem is to find an $u(t)$ that maximizes the QFI defined as 
\beq 
\text{QFI}(T) = \text{Tr} \left[ \hat{\rho}(T) \hat{\mathcal{L}}_\omega^2 (T) \right]  \equiv -\mathcal{C}_Q 
\label{eqn:QFI_DM}
\eeq   
at a given evolution time $T$, with the symmetric logarithmic derivative $\hat{\mathcal{L}}_\omega$ 
\beq 
 \partial_\omega \hat{\rho} \equiv \frac{1}{2} \left(
\hat{\mathcal{L}}_\omega  \hat{\rho}
+ \hat{\rho} \hat{\mathcal{L}}_\omega \right).
\label{eqn:sym_log_der}
\eeq 
In Eq.~\eqref{eqn:QFI_DM}, the negative of QFI is used as the terminal cost function $\mathcal{C}_Q$ to {\em minimize}. In the diagonal basis of $\hat{\rho}$ where $\hat{\rho} | i \rangle = \lambda_i |i\rangle$, $\hat{\mathcal{L}}_\omega = \sum'_{i,j} \frac{2 \langle i | \partial_\omega \hat{\rho} |j \rangle}{\lambda_i + \lambda_j} $ where the $\sum'$ is over non-zero $\lambda_i + \lambda_j$. 

The twist and turn problem without quantum decoherence is studied extensively  \cite{RevModPhys.90.035005} and two important limits are provided. If the initial state is the spin coherent state, then QFI = QFI$_\text{SQL} = N t^2$ is referred to as the standard quantum limit (SQL) which is the largest QFI without entanglement. If the initial state is a Heisenberg-limit (HL) state (also known as GHZ state) where 
\begin{equation}
| \psi_\text{HL} \rangle = \frac{1}{ \sqrt{2} }\left[ 
\big|m_z = \frac{N}{2} \big\rangle + \big|m_z =-\frac{N}{2} \big\rangle \right], 
\label{eqn:HL}
\end{equation}
then QFI = QFI$_\text{HL} = N^2 t^2$ which is the largest QFI one can get \cite{PhysRevLett.96.010401}. Our problem setup follows Ref.~\cite{PhysRevLett.124.060402} where the time required to reach an entangled state is counted as the cost. As explicitly shown in Ref.~\cite{PhysRevA.103.052607}, in the large-$\chi$ limit the optimal control steers the system to HL-state and then stops; the resulting QFI is approaching QFI$_\text{HL}$ upon increasing the evolution time. 
In the presence of decoherence, there appears no general statements about the QFI bound to our knowledge but a super-SQL $N^{4/3}$ scaling is found using continuous nondemolition measurement \cite{PhysRevLett.125.200505}. PMP formalism cannot guarantee the global maximum QFI, but it does provide statements about the optimal control protocol.


\subsection{Decoherence channels and decoherence-free subspace} 

With Eq.~\eqref{eqn:tt_Hamil} represented by $N$ interacting spins, the quantum decoherence is described by
\beq 
\left\{ \frac{d \hat{\rho} }{dt} \right\}_\text{channel}  = \gamma \sum_i \left( \hat{L}_i \hat{\rho} \hat{L}_i^\dagger - \frac{1}{2} \{ \hat{L}_i^\dagger \hat{L}_i, \hat{\rho} \}  \right). 
\label{eqn:drho_short}
\eeq 
$\hat{L}$ is the Lindblad operator whose explicit form specifies the decoherence channel. The decoherence on each spin is assumed to be identical in the $N$-spin system.  We use $i$ to label the spin and consider three standard decoherence channels: the  depolarization channel where
\beq 
\left\{ \frac{d \hat{\rho} }{dt} \right\}_{ \sigma_0 }   = \frac{\gamma}{3} \sum_{i=1}^N \sum_{\alpha=x,y,z} \left( \hat{L}_{i, \alpha} \hat{\rho} \hat{L}_{i, \alpha}^\dagger - \frac{1}{2} \{ \hat{L}_{i, \alpha}^\dagger \hat{L}_{i, \alpha}, \hat{\rho} \}  \right)_{ \hat{L}_{i, \alpha} = \frac{ \sigma_{\alpha}^{(i)} }{2} },
\label{eqn:drho_0}
\eeq 
the dephasing channel where
\beq 
\left\{ \frac{d \hat{\rho} }{dt} \right\}_{ \sigma_z }   = \gamma \sum_{i=1}^N  \left( \hat{L}_{i} \hat{\rho} \hat{L}_{i}^\dagger - \frac{1}{2} \{ \hat{L}_{i}^\dagger \hat{L}_{i}, \hat{\rho} \}  \right)_{\hat{L}_{i} = \frac{ \sigma_{z}^{(i)} }{2} },
\label{eqn:drho_z}
\eeq 
and  the flipping channel 
\beq 
\left\{ \frac{d \hat{\rho} }{dt} \right\}_{ \sigma_x }   = \gamma \sum_{i=1}^N  \left( \hat{L}_{i} \hat{\rho} \hat{L}_{i}^\dagger - \frac{1}{2} \{ \hat{L}_{i}^\dagger \hat{L}_{i}, \hat{\rho} \}  \right)_{ \hat{L}_{i} = \frac{ \sigma_{x}^{(i)} }{2} }. 
\label{eqn:drho_x}
\eeq 
In Eqs.~\eqref{eqn:drho_0} to \eqref{eqn:drho_x}, $\sigma^{(i)}_{\alpha}$ acts only on the $i$th spin and should be formally written as a tensor product $\sigma^{(i)}_{\alpha} \otimes (\Pi_{j \neq i} \otimes \sigma^{(j)}_0)$ with $\sigma_0 \equiv \mathbb{I}$ the 2$\times$2 identity matrix.  
All eigenvalues of $\left\{ \frac{d \hat{\rho} }{dt} \right\}_\text{channel}$ are non-positive; the subspace of zero eigenvalue, excluding the identity $\Pi_i \otimes \sigma_0$, is referred to as the decoherence-free (DF) subspace because any state within this subspace does not dissipate. DF space is channel dependent and is the only allowed non-unity component of $\hat{\rho}$ in the long-time limit. Given that the unity component of $\hat{\rho}$ has zero sensing capability, the DF subspace plays a very crucial role in the asymptotic non-zero QFI; this will be concretely seen in Section \ref{sec:2-spin}.

Because neither the dynamics nor the initial state distinguishes individual spins, the DM only contains the symmetric permutation of tensor-products. We introduce the  ``bar'' notation to represent the summation of all permutations. For example, 
\beq 
\overline{ \sigma_x (\otimes \sigma_z)^2 } = 
\sigma_x \otimes \sigma_z \otimes \sigma_z + 
\sigma_z \otimes \sigma_x \otimes \sigma_z +
\sigma_z \otimes \sigma_z \otimes \sigma_x
\eeq 
This notation will be used for the rest of discussion. 

\subsection{Dynamics and important quantities from PMP} 


In this subsection we summarize expressions relevant for the metrology application, and details of applying PMP to quantum problems can be found in many references \cite{PRXQuantum.2.010101, PRXQuantum.2.030203, PhysRevA.97.062343, PhysRevA.103.052607}. Following Ref.~\cite{PhysRevA.103.052607}, we regard $\hat{\rho}$ and $\partial_\omega \hat{\rho} \equiv \hat{\rho}_\omega$ as {\em independent} dynamical variables. The dynamics for the augmented system composed of $(\hat{\rho}, \hat{\rho}_\omega)$ is 
\beq 
\begin{cases}
 \frac{\partial}{\partial t} \hat{\rho} = -i [\hat{H}_0 + u(t) \hat{H}_1, \hat{\rho}] + \left\{ \frac{d \hat{\rho} }{dt} \right\}, \\ 
 \frac{\partial}{\partial t} \hat{\rho}_\omega = -i [\hat{H}_0 + u(t) \hat{H}_1, \hat{\rho}_\omega] -i [\partial_\omega \hat{H}, \hat{\rho}] + \left\{ \frac{d \hat{\rho}_\omega }{dt} \right\},
\end{cases}
\label{eqn:DM_Lindblad}
\eeq 
where the initial conditions are $\hat{\rho}(0) = | \Psi_\text{coh-x} \rangle \langle \Psi_\text{coh-x} | $ and $\hat{\rho}_\omega (0) = 0$. We note that $\hat{\rho}$ does not depend on $\hat{\rho}_\omega$; $-i [\partial_\omega \hat{H}, \hat{\rho}]$ is the source term of $\hat{\rho}_\omega$. 
In optimal control theory, the control system described by Eqs.~\eqref{eqn:DM_Lindblad} is referred to be ``time-invariant'' because the time dependence of dynamics is exclusively through the control $u(t)$; it is referred to be ``control-affine'' because the dynamics depends linearly on $u(t)$.

For a given dynamical system, PMP introduces an auxiliary problem for costate variables that satisfy similar dynamics for the original dynamical variables; by doing so the necessary conditions for an optimal solution can be compactly expressed and thus efficiently computed. With the augmented dynamics Eqs.~\eqref{eqn:DM_Lindblad}, we denote the corresponding costate variables as $\hat{\lambda} \leftrightarrow \hat{\rho}$ and $ \hat{\lambda}_\omega \leftrightarrow \hat{\rho}_\omega $. The control-Hamiltonian (c-Hamiltonian, $\mathcal{H}_{oc}$) is defined as 
\beq 
\begin{aligned}
\mathcal{H}_{oc} &= \text{Tr} \left\{ \hat{\lambda} ( \partial_t \hat{ \rho } ) \right\} + \text{Tr} \left\{ \hat{\lambda}_\omega ( \partial_t \hat{ \rho }_\omega ) \right\} \sim \frac{\partial \mathcal{C}_Q}{\partial T},
\end{aligned} 
\label{eqn:Hoc_DM}
\eeq 
which is a scalar corresponding to the time derivative of the terminal cost (negative of QFI). Negative/positive $\mathcal{H}_{oc}$ thus indicates an increase/decrease of QFI$(T)$ upon increasing $T$. The switching function $\Phi$  is 
\beq 
\begin{aligned}
\Phi(t) &= 
-i \text{Tr} \left\{ \hat{\lambda}  [\hat{H}_1, \hat{\rho}] \right\} -i \text{Tr} \left\{ \hat{\lambda}_\omega  [ \hat{H}_1, \hat{\rho}_\omega]  \right\} 
\sim \frac{\partial \mathcal{C}_Q }{\partial u(t)},
\end{aligned} 
\label{eqn:Phi_DM}
\eeq 
which corresponds to the gradient the terminal cost function with respect to the control and has been used to obtain the numerical solution in the gradient-based optimization algorithm \cite{PhysRevA.103.052607}. 
The costate equations of motion obtained by $\frac{\partial}{\partial t} \hat{\lambda} = -\frac{\partial \mathcal{H}_{oc} }{\partial \hat{\rho} }$ and $\frac{\partial}{\partial t} \hat{\lambda}_\omega = -\frac{\partial \mathcal{H}_{oc} }{\partial \hat{\rho}_\omega }$ are 
\beq 
\begin{cases}
 \frac{\partial}{\partial t} \hat{\lambda} = -i [\hat{H}_0 + u(t) \hat{H}_1, \hat{\lambda}]  - i [ \partial_\omega \hat{H}, \hat{\lambda}_\omega] - \left\{ \frac{d \hat{\lambda} }{dt} \right\}, \\
 \frac{\partial}{\partial t} \hat{\lambda}_\omega = -i [\hat{H}_0 + u(t) \hat{H}_1,  \hat{\lambda}_\omega] - \left\{ \frac{d \hat{\lambda}_\omega }{dt} \right\}.
\end{cases}
\label{eqn:DM_QFI}
\eeq 
We have used $\hat{L}_i = \hat{L}_i^\dagger$ in the derivation. 
The boundary conditions for costate variables $\hat{\lambda}$ and $\hat{\lambda}_\omega$ and can be symbolically described as $\hat{\lambda}(T) = + \frac{\partial \mathcal{C}_{Q} }{ \partial \hat{\rho}^\dagger }$, $\hat{\lambda}_\omega (T) = + \frac{\partial \mathcal{C}_{Q} }{ \partial \hat{\rho}_\omega^\dagger }$.
Due to its practical importance the derivations will be provided shortly in Section \ref{sec:lambda_boundary}. Once $\hat{\rho}(t)$, $\hat{\rho}_\omega(t)$, $\hat{\lambda} (t)$, and $\hat{\lambda}_\omega (t)$ are solved, $\mathcal{H}_{oc}(t)$ and $\Phi(t)$ can be evaluated. It is worth mentioning that while PMP is designed for real-valued dynamical variables \cite{book:GeometricOptimalControl}, the formalism provided here preserves the dynamics in its natural Lindblad form with complex-valued dynamical and costate variables (i.e., $\hat{\rho}$ and $\hat{\lambda}$) and at the same time guarantees a real-valued switching function and c-Hamiltonian.

According to PMP, the first-order necessary conditions for an optimal control $u^*(t)$ are such that 
\begin{subequations}
\begin{align}
u^*(t) &= \begin{cases} 
|u_\text{max} | & \Phi(t) < 0 \\
-|u_\text{max} | & \Phi(t) > 0 \\ 
\text{undetermined} & \Phi(t) = 0.           
\end{cases},  
\label{eqn:Phi_condition}    \\
\mathcal{H}_{oc}(t) &= \text{const.} \label{eqn:H_condition}
\end{align}
\label{eqn:1st_order_PMP}
\end{subequations}
Conditions \eqref{eqn:1st_order_PMP} can be used to quantify the quality of a solution. When $\Phi \neq 0$, controls take the extreme values and are referred to as ``bang'' control. When $\Phi = 0$ over a finite range of time, the control values cannot be determined from the first-order condition; the resulting optimal control is referred to as ``singular'' control. 

Assuming the control is singular over a finite time interval, then all time derivatives of $\Phi$ have to vanish, including the second derivative
\beq 
\ddot{\Phi} \equiv u(t) \, \langle [g, [f,g]] \rangle(t) +  \langle [f, [f,g]] \rangle(t).
\label{eqn:ffg}
\eeq 
Here $\langle [g, [f,g]] \rangle(t)$ and $\langle [f, [f,g]] \rangle(t)$ simply represent two real-valued functions of time which can be straightforwardly computed for a given control $u(t)$ \cite{ffg-notation}. Expressions for flipping channel will be provided in the Appendix A [Eq.~\eqref{eqn:u_singular_flipping}]. 
According to PMP, a singular control requires 
\begin{subequations}
\begin{align}
&u_\text{sing}(t) = -\frac{ \langle [f, [f,g]] \rangle(t) }{  \langle [g, [f,g]] \rangle(t) }, 
\label{eqn:2nd_u*} \\
&\langle [g, [f,g]] \rangle(t) \leq 0. \label{eqn:LC_condition}
\end{align}
\label{eqn:2nd_order}
\end{subequations} 
Eq.~\eqref{eqn:LC_condition} is known as the Legendre-Clebsch condition and is the second-order necessary condition for optimal controls (Chapter 4 of Ref.~\cite{book:GeometricOptimalControl}). If gradient-based method finds the vanishing $\Phi(t)$ over a time interval, the corresponding control should be numerically close to Eq.~\eqref{eqn:2nd_u*}. Moreover, if we know that the optimal control is singular, Eq.~\eqref{eqn:2nd_u*} provides a self-consistency equation for determining $u^*(t)$. Eq.~\eqref{eqn:2nd_u*} has been applied to the dissipative qubit system \cite{PhysRevA.101.022320} and continuous-time quantum computation \cite{PhysRevLett.126.070505}. Eq.~\eqref{eqn:LC_condition} is seldom applied to any realistic control systems due to its complexity but we shall use it to locate the onset of instability beyond which bang controls are intrinsically needed for a bounded solution [see Section \ref{sec:sigma_x}].

\subsection{Costate boundary condition \label{sec:lambda_boundary}} 

To avoid ambiguities, we will write the costate boundary conditions (at the evolution time $T$) in component form. The symmetric logarithm derivative $\hat{\mathcal{L}}_\omega$ is defined as 
\beq 
2 \rho_{\omega, ij} = \sum_k \left( \rho_{ik} \mathcal{L}_{\omega, kj} + \mathcal{L}_{\omega, ik} \rho_{ kj} \right).
\eeq 
We will need $\frac{\delta \mathcal{L}_{\omega, ij} }{\delta \rho_{ab}}$ and $\frac{\delta \mathcal{L}_{\omega, ij} }{\delta \rho_{\omega, ab} }$ which in component form include four indices.  
\beq 
\begin{aligned}
2 \frac{\partial \rho_{\omega,ij}}{ \partial \rho_{ab} } 
&= 0 = \sum_k \left( \rho_{ik} \frac{\partial \mathcal{L}_{\omega, kj} }{  \partial \rho_{ab} } + \frac{\partial \mathcal{L}_{\omega, ik} }{  \partial \rho_{ab} } \rho_{kj} \right)
+ \left( \delta_{ia}  \mathcal{L}_{\omega, bj} + \mathcal{L}_{\omega, ia} \delta_{bj}   \right), \\
2 \frac{\partial \rho_{\omega,ij}}{ \partial \rho_{\omega,ab} } 
&= 2 \delta_{ia} \delta_{jb} = \sum_k \left( \rho_{ik} \frac{\partial \mathcal{L}_{\omega, kj} }{  \partial \rho_{\omega,ab} } + \frac{\partial \mathcal{L}_{\omega, ik} }{  \partial \rho_{\omega,ab} } \rho_{kj} \right). 
\end{aligned}
\label{eqn:SLD_derivative_01}
\eeq 
Eqs.~\eqref{eqn:SLD_derivative_01} allow us to solve $\frac{\delta \mathcal{L}_{\omega, ij} }{\delta \rho_{ab}}$ and $\frac{\delta \mathcal{L}_{\omega, ij} }{\delta \rho_{\omega, ab} }$; the easiest way is to solve them in the eigenbasis of $\hat{\rho}$ and then transform them back [see the expression below Eq.~\eqref{eqn:sym_log_der}]. As $\rho_{ij}$ and $\rho_{\omega, ij}$ are independent and their partial derivatives vanish. Also, we solve  $\frac{\delta \mathcal{L}_{\omega, ij} }{\delta \rho_{ab} }$ for all $i,j$ and given $a,b$; similarly  for $\frac{\delta \mathcal{L}_{\omega, ij} }{\delta \rho_{\omega, ab} }$. Although $\frac{\delta \mathcal{L}_{\omega, ij} }{\delta \rho_{ab} }$ has $N^4$ components (each $i,j,a,b$ can go from 1 to $N$) and in principle requires $N^4$ linear equations, one practically decomposes them into $N \times N$ decoupled sub-problems, with each sub-problem having $N \times N$ variables.  

Defining the terminal cost function as the negative of QFI, i.e., $\mathcal{C}_Q =  -\sum_{ij} \rho_{ij} \left( \hat{\mathcal{L}}^2_\omega \right)_{ji}$ we get the boundary conditions for costates as 
\beq 
\begin{aligned}
\lambda_{ba} &= \lambda^*_{ab} = -\frac{\partial }{\partial \rho_{ab} } \text{Tr}(\hat{\rho} \hat{\mathcal{L}}_\omega^2) = -\left[ \left(\hat{\mathcal{L}}_\omega^2\right)_{ba} + \sum_{ijk} \rho_{ij}\left(   \frac{\delta \mathcal{L}_{\omega, jk} }{\delta \rho_{ab}} \mathcal{L}_{\omega, ki} +  \mathcal{L}_{\omega, jk} \frac{\delta \mathcal{L}_{\omega, ki} }{\delta \rho_{ab} }  \right) \right], \\
\lambda_{\omega, ba} &= \lambda_{\omega, ab}^* = -\frac{\partial }{\partial \rho_{\omega, ab} } \text{Tr}(\hat{\rho} \hat{\mathcal{L}}_\omega^2) = -\left[  \sum_{ijk} \rho_{ ij}\left(   \frac{\delta \mathcal{L}_{\omega, jk} }{\delta \rho_{\omega, ab}} \mathcal{L}_{\omega, ki} +  \mathcal{L}_{\omega, jk} \frac{\delta \mathcal{L}_{\omega, ki} }{\delta \rho_{\omega, ab} }  \right) \right].
\end{aligned}
\label{eqn:costate_boundary_DM}
\eeq 
Beware of the {\em transpose} relationship between the costate matrix and the partial derivative that takes the complex-valued dynamical variables into account. Expressions for classical Fisher information are provided in Appendix B.

\subsection{Short summary and advantages of augmentation}
\begin{table}[ht]
\begin{tabular}{l| ll} 
classification & symbol & description \\ \hline 
quantities & $\hat{\rho}$ & density matrix, main dynamical variables  \\ 
from  & $\hat{\rho}_\omega$& $\partial_\omega \hat{\rho}$, complex-valued matrix,
augmented dynamical variable  \\ 
original & $\hat{\mathcal{L}}_\omega$& $\hat{\rho}_\omega = \frac{1}{2} \{ \hat{\rho}, \hat{\mathcal{L}}_\omega\} $
symmetric logarithm derivative  \\ 
problem & QFI & $\text{Tr} (\hat{\rho} \hat{\mathcal{L}}^2_\omega)$, quantum Fisher information, terminal cost to maximize
\\ \hline 
quantities & $\hat{\lambda}$ &  complex-valued matrix, costate variable  \\  
introduced & $\hat{\lambda}_\omega$& $\partial_\omega \hat{\lambda}$, complex-valued matrix,
augmented costate variable  \\ 
by PMP & $\Phi$ & switching function, a real-valued scalar, 0 for singular control \\ 
 & $\mathcal{H}_{oc}$ & control Hamiltonian, a real-valued scalar, constant for optimal solution \\ 
 & $\langle [g, [f,g]] \rangle$ & $\ddot{\Phi} \equiv \langle [g, [f,g]] \rangle u(t) + \langle [f, [f,g]] \rangle$, real-valued scalar, negative for singular control \\ \hline
dynamics & $\hat{H}(t)$ & Hamiltonian, complex valued matrix describing the unitary dynamics \\
specification & $\{ \frac{d\hat{\rho} }{dt}\}_\text{channel} $ &
 operator describing quantum decoherence, subscript indicating dissipation channel
\end{tabular}
\caption{Important quantities and their symbols. $\omega$ is the external parameter to be estimated. 
}
\label{table:definition}
\end{table} 

As the notations are a bit involved, in Table \ref{table:definition} we summarize the main variables and their primary roles. We distinguish the variables of the original dynamics (the density matrix and its derivatives) from those introduced by PMP. In addition to costate variables, PMP introduces at least three relevant scalar functions. The switching function $\Phi$ serves because the gradient of terminal cost with respect to control $u(t)$ and is the most crucial practical step in numerical optimization. The gradient-based optimization procedure reads
\beq 
u^{(n+1)}(t) \leftarrow u^{(n)}(t) - \text{learning rate} \times \Phi(t). 
\label{eqn:updating_u}
\eeq 
The condition \eqref{eqn:Phi_condition} is intuitive as optimality implies a vanishing gradient. 
The c-Hamiltonian $\mathcal{H}_{oc}$ corresponds to $\frac{\partial \mathcal{C}_Q}{\partial T}$ and is a constant during the entire evolution for an optimal solution [Eq.~\eqref{eqn:H_condition}]. The latter may not be intuitive and the flatness of $\mathcal{H}_{oc}(t)$ can be an indicator of the solution quality no matter the controls are bang or singular (recall bang control implies non-zero $\Phi$). Finally, $\langle [g, [f,g]] \rangle$ and $\langle [f, [f,g]] \rangle$ are relevant for singular controls, which seem prevalent in quantum problems \cite{PhysRevLett.111.260501, PhysRevA.101.022320, PhysRevA.103.052607}. In particular, Eq.~\eqref{eqn:2nd_u*} directly gives the values on the singular controls and can be used in determining optimal solutions. 


The augmented dynamics \eqref{eqn:DM_Lindblad} is designed to keep the resulting control problem {\em time-invariant} and we now elaborate its advantages in both obtaining the solution and quantifying the solution quality. 
For obtaining the solution, it greatly reduces the computational cost of evaluating the gradient $\Phi(t) \sim \frac{\delta \mathcal{C}}{\delta u(t)}$ as Eq.~\eqref{eqn:Phi_DM} only requires $u(t)$ at a single time point $t$. In contrast, evaluating $\frac{\delta \mathcal{C}}{\delta u(t)}$ using the straightforward GRAPE algorithm \cite{Khaneja-2005, PhysRevA.96.012117} would need the history of $u(t')$ with $0\leq t' \leq t$ (see the Appendix in Ref.~\cite{PhysRevA.96.012117}). Certainly once a proper augmented dynamics is identified, the GRAPE algorithm and PMP are equivalent in computing the gradient. In this sense what PMP provides is the criterion (i.e., to make the control problem time-invariant) of the suitable augmented dynamics which is terminal-cost specific. 
Once the control system is kept time-invariant and control-affine, PMP offers two additional optimality conditions, i.e., Eq.~\eqref{eqn:H_condition} and Eqs.~\eqref{eqn:2nd_order}, that are beyond the gradient and can be used to further constrain the optimal solution. For example the flatness of c-Hamiltonian has been used to quantify the discretization error  \cite{PhysRevA.103.052607}. As will be shown in Section \ref{sec:2-spin}, applying the second-order conditions [Eqs.~\eqref{eqn:2nd_order}] reveals that the optimal control has to contain the bang segments to be bounded in some dissipation channel.

\section{Results for 2-spin systems \label{sec:2-spin}}  

\subsection{Overview and depolarization channel}
\begin{figure}[ht]
\centering
\includegraphics[width=0.48\textwidth]{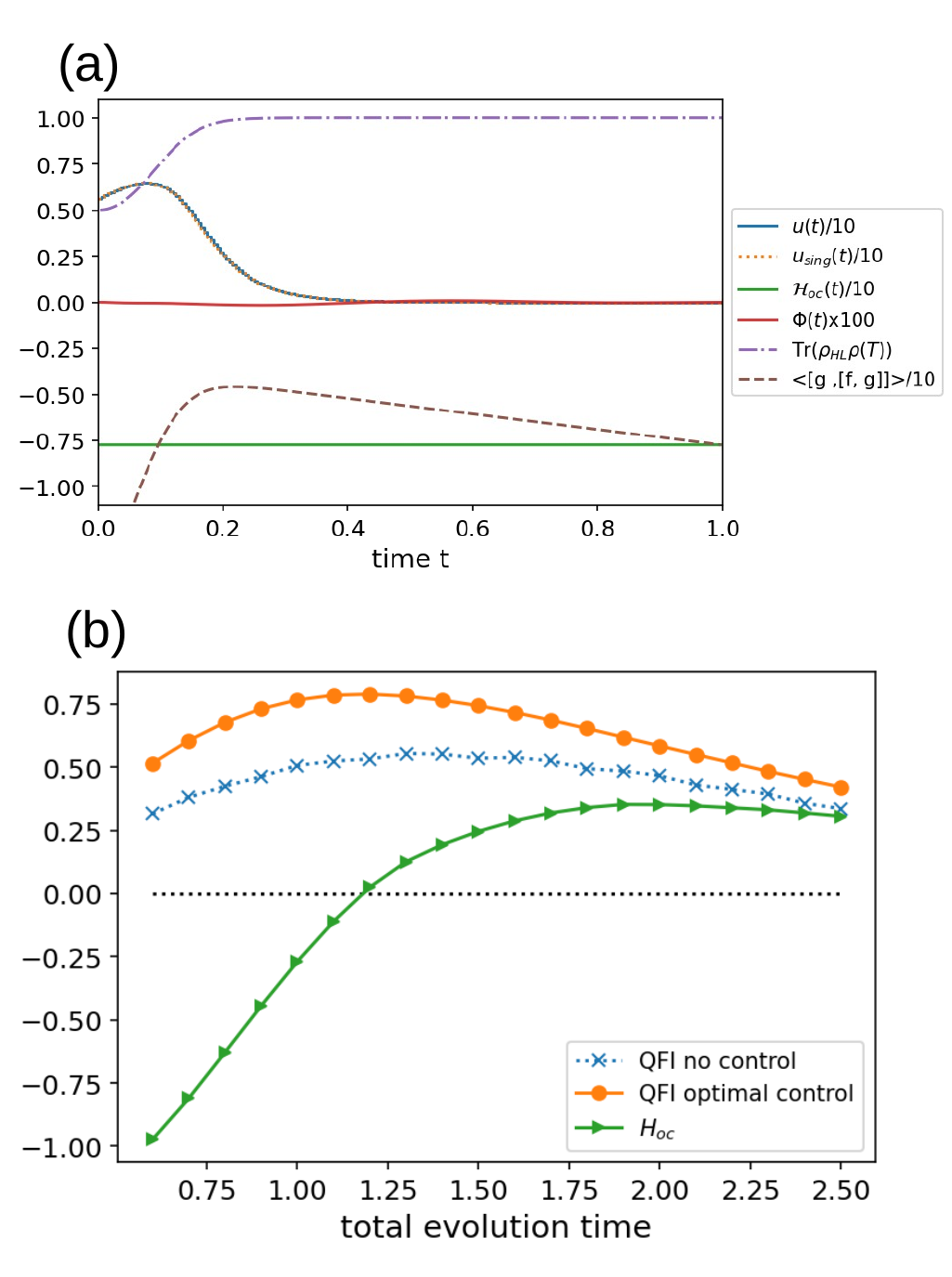}
\caption{ (a) Reference problem for 2-spin, $\chi=10$, $T=1$. All necessary conditions are satisfied: $|\Phi(t)|$ is small (note its values are multiplied by 100) in the plot; $\mathcal{H}_{oc}(t)$ is flat; $\langle [g,[f,g]]\rangle$ is negative; the optimal control based on gradient is close to that obtained from Eq.~\eqref{eqn:2nd_u*}. $\text{Tr}[\hat{\rho}_\text{HL} \hat{ \rho}(t)]$ approaches unity around $t=0.2$ indicating that the system is steered to the HL-state. 
(b) The QFI for depolarization channel as a function of evolution time. QFI with control is larger than that without control. The maximum QFI$(T)$ corresponds to vanishing $\mathcal{H}_{oc}$.
}
\label{fig:N2_Reference_G1p5_Lxyz}
\end{figure}

In this section we provide detailed results for two-spin systems under three dissipation channels. We focus on the strong-coupling limit where the dimensionless parameter $N \chi T \gg 1$ ($\chi$ has the dimension of energy and $\hbar \equiv 1$ is used); specific parameters used in this section are $N=2$, $\chi=10$, $\omega=0$, $\gamma=1.5$. The choice of $\gamma$, which has the dimension of 1/time, is such that the decoherence effect can be observed around $T=1$-4. As a reference, Fig.~\ref{fig:N2_Reference_G1p5_Lxyz}(a) gives the optimal control without decoherence ($\gamma=0$) for $T=1$. All first-order and second-order necessary conditions are numerically satisfied (see figure caption for details). Without decoherence, the control in this limit brings the system quickly to the HL state [Eq.~\eqref{eqn:HL}] during the early evolution and stops \cite{PhysRevA.103.052607}. This is explicitly shown  in Fig.~\ref{fig:N2_Reference_G1p5_Lxyz}(a) as $\text{Tr}[\hat{\rho}_\text{HL} \hat{ \rho}(t)]$ approaches unity around $t=0.2$; here $\hat{\rho}_\text{HL} $ is the DM corresponding to 2-spin HL state:
\beq 
\hat{\rho}_\text{HL} 
= \frac{1}{4} \left[ \sigma_0 \otimes \sigma_0 
+ \sigma_x \otimes \sigma_x 
- \sigma_y \otimes \sigma_y 
+ \sigma_z \otimes \sigma_z \right]. 
\label{eqn:rho_HL}
\eeq

With decoherence, we shall use the DF subspace and the HL-state to understand 
some qualitative behavior. 
For the depolarization channel, there is no DF subspace so $\hat{\rho} \rightarrow \frac{1}{4} \sigma_0 \otimes \sigma_0 $ and $\hat{\rho}_\omega \rightarrow 0$ in the long-time limit. We expect and indeed see that the QFI$(t)$ is peaked at a finite $T$ and decays to zero as shown in Fig.~\ref{fig:N2_Reference_G1p5_Lxyz}(b). Numerically we explicitly show that QFI with control is larger than that without control, and the optimal evolution time (i.e., maximum QFI$(T)$) occurs at $\mathcal{H}_{oc}=0$.

Both the dephasing and flipping channels have non-zero DF subspace so the asymptotic QFI can be non-zero. Let us discuss the qualitative impact of a decoherence channel by considering its effect on the HL state -- the best sensing state without decoherence. Recall that the 2-spin HL-state for is $\frac{1}{\sqrt{2}} (| \uparrow \uparrow \rangle + | \downarrow \downarrow \rangle)$. A decoherence channel can be regarded as an unwanted measurement done by the environment. The dephasing channel projects the local spin to the eigenbasis of $\sigma_z$. As a result one measurement collapses the HL-state to either $| \uparrow \uparrow \rangle$ or $| \downarrow \downarrow \rangle$ which is completely insensitive to $\hat{J}_z$. The flipping channel projects the local spin to the eigenbasis of $\sigma_x$. As a result one measurement collapses the HL-state to  $\frac{1}{2}( | \uparrow \rangle \pm | \downarrow \rangle ) \otimes ( | \uparrow \rangle \pm | \downarrow \rangle )$  which is still sensitive to $\hat{J}_z$. 
We thus expect that the QFI under flipping channel is larger than that under the dephasing channel. As will be shown in following subsections, simulations are consistent with this expectation.

\subsection{Dephasing channel}

\begin{figure}[ht]
\centering
\includegraphics[width=0.48\textwidth]{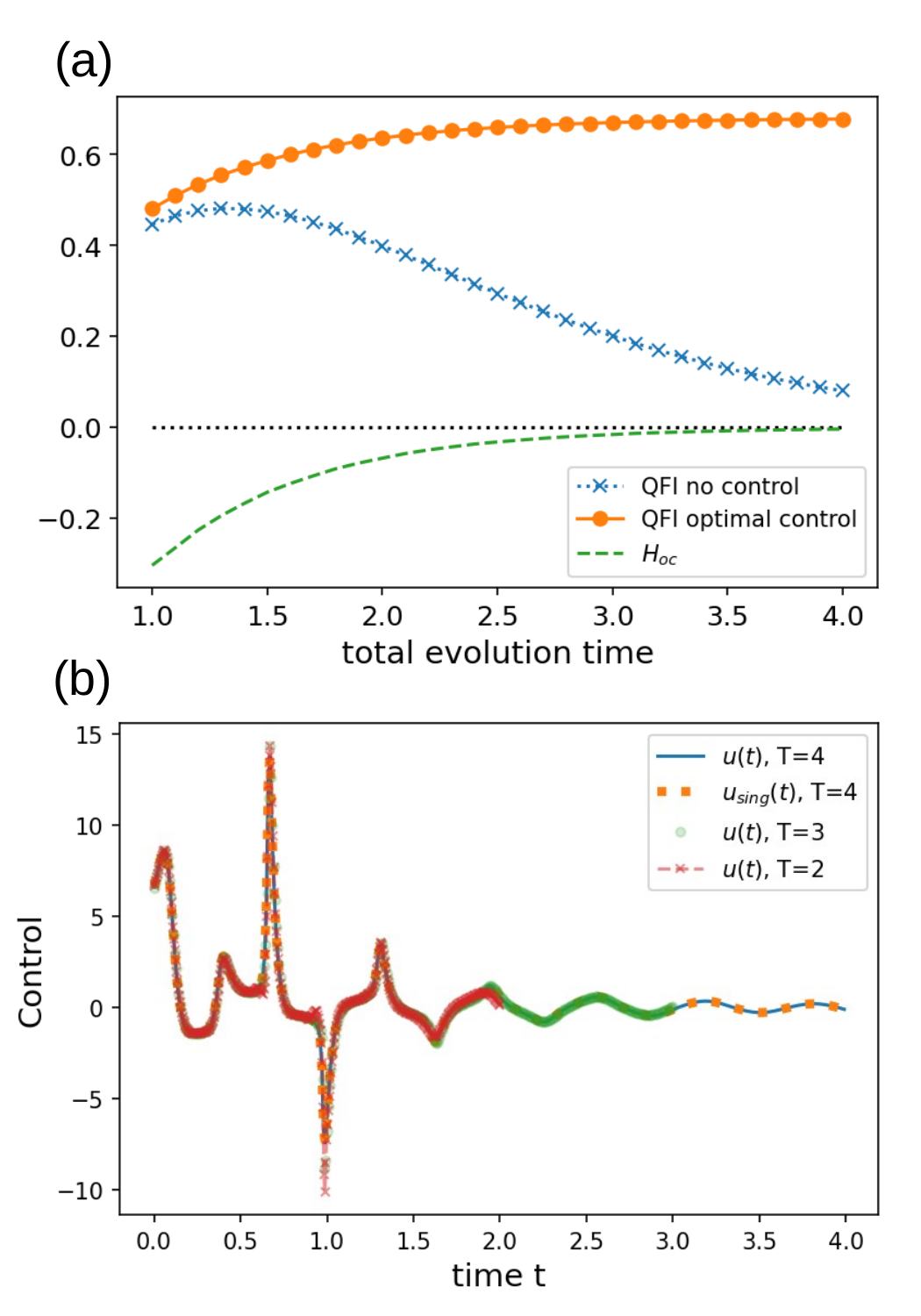}
\caption{ (a) The QFI for dephasing channel as a function of evolution time: it saturates to a non-zero value in the long-time limit. QFI with control is larger than that without control.  The maximum QFI$(T)$ corresponds to vanishing $\mathcal{H}_{oc}$. (b) Obtained optimal controls for $T=4,3,2$. For $T=4$, the singular control using  Eq.~\eqref{eqn:2nd_u*} is also given and it practically coincides with that obtained by the gradient based method. 
}
\label{fig:N2_G1p5_Lz}
\end{figure}  

For the dephasing channel, the non-trivial (excluding the identity matrix) DF subspace is $\sigma_z \otimes \sigma_z$ and $\overline{\sigma_0 \otimes \sigma_z}$. Given that both $-i [\hat{J}_z, \sigma_z \otimes \sigma_z] = -i [\hat{J}_z, \overline{\sigma_0 \otimes \sigma_z}] = 0$, when $\hat{\rho}$ is in the DF-subspace it cannot be the source for $\hat{\rho}_\omega$. In the long-time limit, both $\hat{\rho}$ and $\hat{\rho}_\omega$ are in DF subspace and we have 
\beq 
\hat{\rho} = \frac{1}{4} \left[ \sigma_0 \otimes \sigma_0 + a_z \sigma_z \otimes \sigma_z\right], \,\,
\hat{\rho}_\omega = b_z \overline{\sigma_0 \otimes \sigma_z}.
\label{eqn:asym_sigma_z}
\eeq  
Using Eq.~\eqref{eqn:sym_log_der} $\hat{\mathcal{L} }_\omega = \frac{4 b_z}{1 + a_z} \overline{\sigma_0 \otimes \sigma_z}$ and the asymptotic QFI is $\frac{32 b_z^2}{1+a_z} >0$. The values of $a_z$, $b_z$ depend on the control and have to be determined numerically. We also note that QFI cannot decrease in the long-time limit because QFI$(T_0 + dt)$ is at least as large as  QFI$(T_0)$ without control and therefore a saturation is expected.

Fig.~\ref{fig:N2_G1p5_Lz}(a) shows QFI$(T)$, and we indeed see that it saturates to a finite value around 0.678  in the long-time limit. Numerically we determine $a_z \approx 0.0786$ and $b_z \approx -0.151$. $\mathcal{H}_{oc}$ approaching zero from the negative side is consistent with the saturation behavior. 
The obtained optimal controls for $T=4,3,2$ are provided in Fig.~\ref{fig:N2_G1p5_Lz}(b). For $T=4$, Eq.~\eqref{eqn:2nd_u*} is used to confirm its optimality. An oscillation is developed to fight the decoherence. We find that there can be several numerical solutions that give similar QFI$(T)$ and small $|\Phi(t)|$, especially when $T$ is long. In this sense the value of optimal QFI$(T)$ is more robust than the numerically obtained optimal control.

\subsection{Flipping channel \label{sec:sigma_x}} 
The flipping channel is the most interesting case because it allows the largest maximum QFI. For the flipping channel, the non-trivial  DF subspace is $\sigma_x \otimes \sigma_x$ and $\overline{\sigma_0 \otimes \sigma_x}$. $[\hat{H}(t), \sigma_x \otimes \sigma_x] = 0$ further indicates that the component of $\sigma_x \otimes \sigma_x$ does not change during the entire evolution. Because $-i [\hat{J}_z, {\sigma_x \otimes \sigma_x}] = \overline{\sigma_x \otimes \sigma_y}$, $\sigma_x \otimes \sigma_x$ component of $\hat{\rho}$  injects $\overline{\sigma_x \otimes \sigma_y}$ to $\hat{\rho}_\omega$ and this can last forever.  Assuming $\hat{\rho}_\omega = b(t) \overline{\sigma_x \otimes \sigma_y}$ and $\hat{\rho} = \frac{1}{4} \left[ \sigma_0 \otimes \sigma_0 + a_x  \sigma_x \otimes \sigma_x\right]$, one gets an equation for $\dot{b} = \frac{ a_x}{4} - \frac{\gamma}{2} b$ whose steady-state solution is $b_\infty = \frac{ a_x}{2\gamma}$, i.e., $\hat{\rho}_\omega = \frac{a_x}{2 \gamma} \overline{\sigma_x \otimes \sigma_y}$ in the long-time limit. $\hat{\mathcal{L} }_\omega = \frac{2 a_x}{\gamma} \overline{\sigma_x \otimes \sigma_y}$ from Eq.~\eqref{eqn:sym_log_der} and the asymptotic QFI is 
\beq 
\text{QFI}_{\infty} = \frac{8 a_x^2}{\gamma^2}
\eeq 
When $| \Psi_\text{coh-x} \rangle$ is the initial state $a_x=1$ and $\text{QFI}_{\infty} = \frac{8 }{\gamma^2}$. For $\gamma=1.5$ $\text{QFI}_{\infty} \approx 3.556$.  We emphasize that the asymptotic value depends {\em only} on the initial state via $a_x$  because the $\sigma_x \otimes \sigma_x$ component does not change under the flipping channel.

\begin{figure}[ht]
\centering
\includegraphics[width=0.48\textwidth]{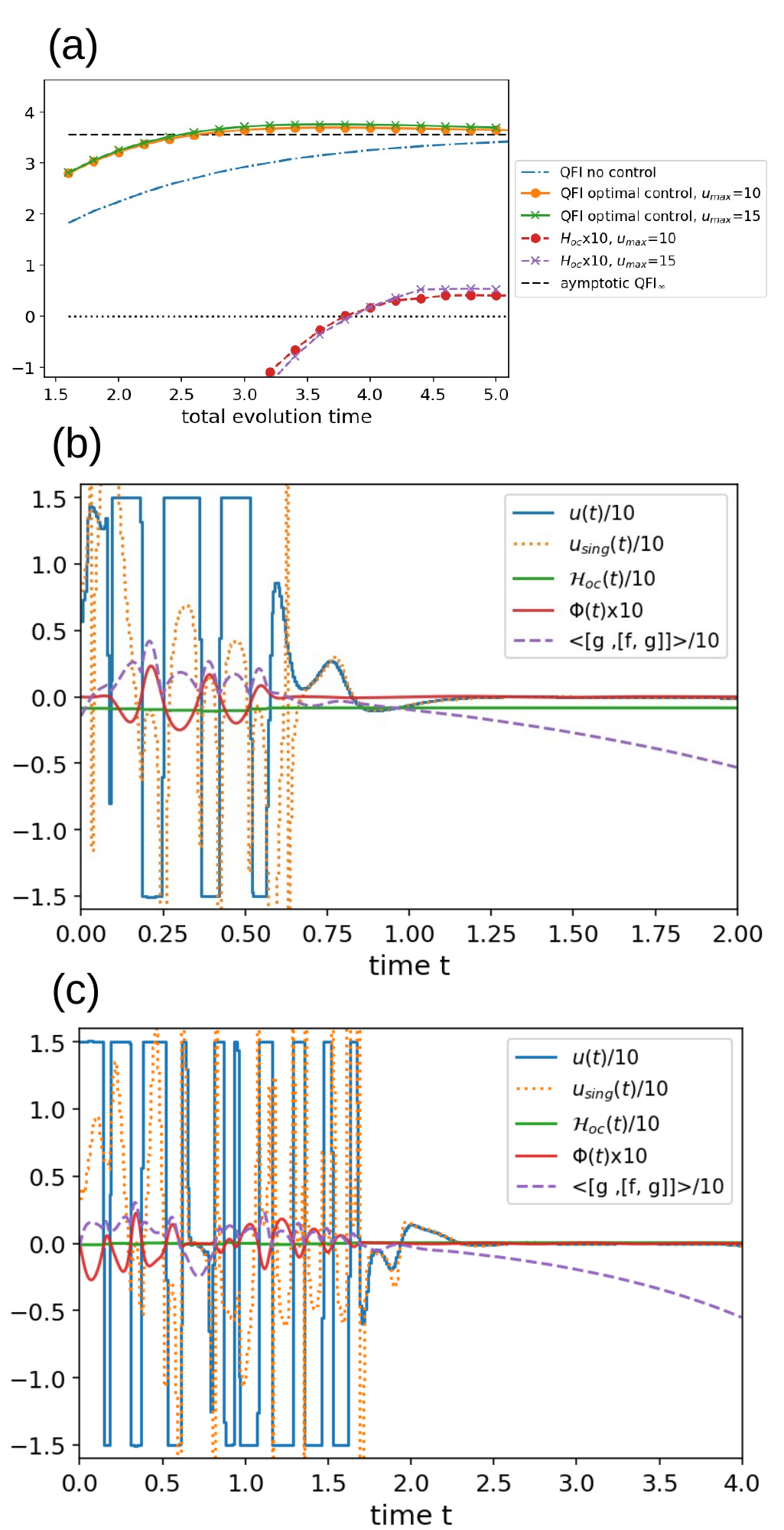}
\caption{ Results of $N=2$, $\chi=10$, $\gamma=1.5$ for flipping channel. 
(a) Optimal QFI with $u_\text{max}=15$ and 10. Applying controls always give a greater QFI, but the difference to QFI's without controls decreases as $T$ increases. Eventually they converge to the same asymptotic $\text{QFI}_\infty = 8/\gamma^2 \approx 3.556$. 
(b) The optimal control for $T=2$ with $u_\text{max}= 15$.
(c) The optimal control for $T=4$ with $u_\text{max}= 15$. 
In panels (b) and (c), the optimal control includes bang and singular parts. The singular part of the control agrees well with the second-order condition Eq.~\eqref{eqn:2nd_u*}. 
}
\label{fig:N2_G1p5_Lx_QFI_control}
\end{figure} 

For reasons which will be presented shortly, the amplitude constraint is needed to have a finite solution; otherwise the controls can become unbounded at some time points. We use two amplitude constraints $u_\text{max}= 10$ and 15, and the resulting QFI$(T)$ are shown in Fig.~\ref{fig:N2_G1p5_Lx_QFI_control}(a). Compared to QFI without controls, the QFI with controls is greater but their difference decreases as $T$ increases because both eventually converge to the same asymptotic $\text{QFI}_\infty = 8/\gamma^2 \approx 3.556$. Fig.~\ref{fig:N2_G1p5_Lx_QFI_control}(b) and (c) present the optimal controls with $u_\text{max}= 15$ which contain both bang and singular parts. The necessary conditions are to a good approximation satisfied. In particular we see that when the control is singular, the values are very close to those computed using Eq.~\eqref{eqn:2nd_u*} and the $\langle [g,[f,g]] \rangle$ is negative, i.e., both second-order conditions \eqref{eqn:2nd_order} are well satisfied. For large $T$ where the decoherence effect dominates, the frequency of switching between bangs increases. This is similar to the dynamical decoupling \cite{PhysRevLett.82.2417} where an oscillation in control can suppress the decoherence effect.

\begin{figure}[ht]
\centering
\includegraphics[width=0.48\textwidth]{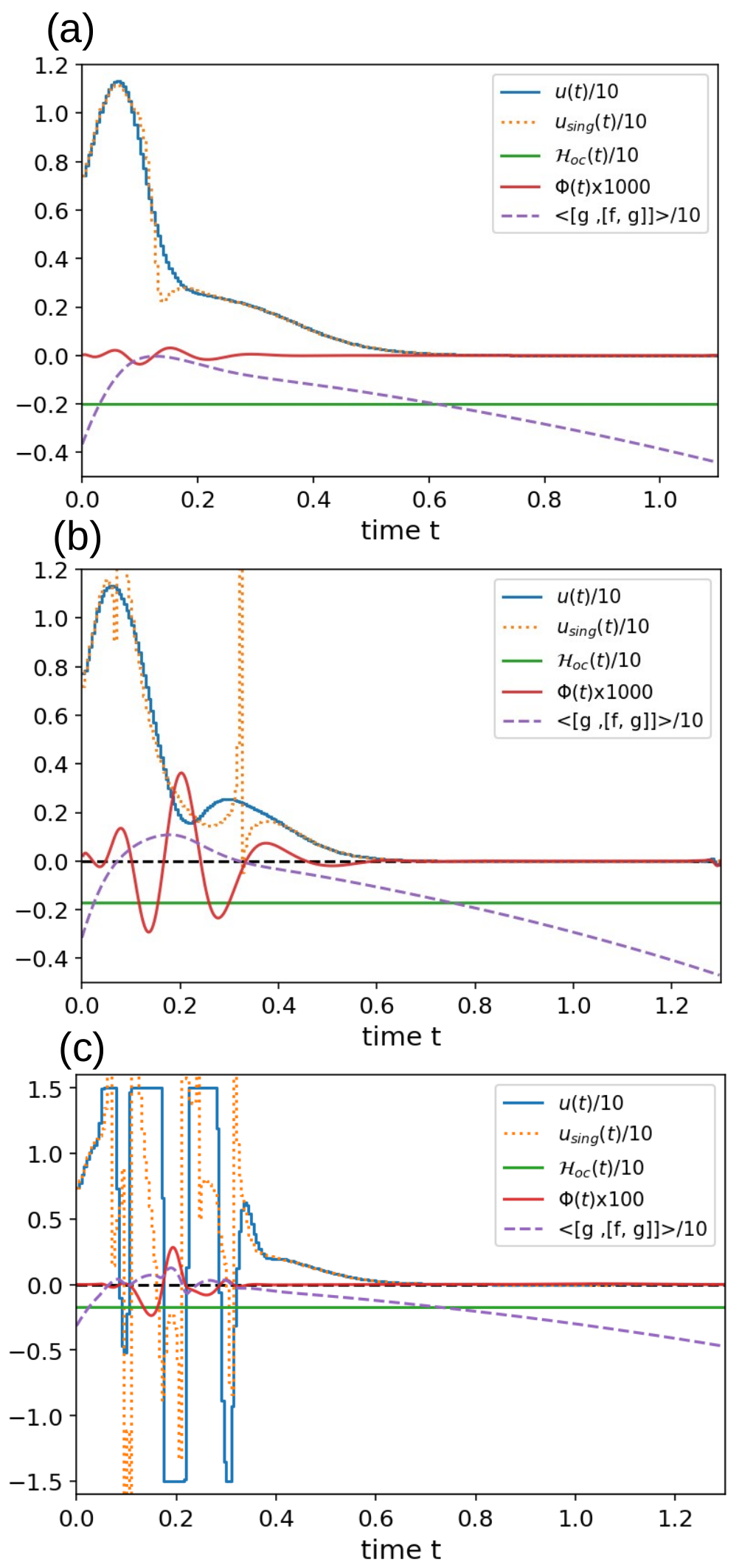}
\caption{ 
Controls near the onset of instability for the flipping channel ($N=2$, $\chi=10$, $\gamma=1.5$). 
(a) The onset of instability occurs at $T=1.1$ where $\langle [g,[f,g]] \rangle$ are mostly negative but approaches zero around $t=0.1$. 
(b) $T=1.3$ assuming a singular control over the entire evolution. The Legendre-Clebsch condition \eqref{eqn:LC_condition} is violated around $t$=0.1 to 0.3. The resulting QFI$(T)$ = 2.338. 
(c) A singular-bang-singular solution with an amplitude constraint $u_\text{max}=15$. QFI$(T)$ = 2.341. Without amplitude constraint, the control becomes unbounded upon iterations of Eq.~\eqref{eqn:updating_u}. 
}
\label{fig:N2_G1p5_T1p3_Lx}
\end{figure} 

We now elaborate the need of an amplitude constraint for a bounded solution. Our numerical simulations suggest that the onset of instability where the control cannot be singular during the entire evolution occurs around $T=1.1$. As shown in Fig.~\ref{fig:N2_G1p5_T1p3_Lx}(a), the indicator of instability is the vanishing $\langle [g,[f,g] ]\rangle$ around $t=0.1$. 
Near the onset of instability at $T=1.3$, one can still numerically get a singular control solution shown in Fig.~\ref{fig:N2_G1p5_T1p3_Lx}(b). There is a finite time interval around 0.1 to 0.3 where the Legendre-Clebsch condition \eqref{eqn:LC_condition} is violated, indicating that the controls can be unbounded and an amplitude constraint is needed. Fig.~\ref{fig:N2_G1p5_T1p3_Lx}(c) shows a solution with an amplitude constraint $u_\text{max} = 15$. Around $t=0.1$ to 0.3 the optimal controls are the bangs. Without the amplitude constraint, the control becomes unbounded upon iterations of Eq.~\eqref{eqn:updating_u}. 
Near the instability, using first-order conditions alone is hard to tell if a solution is a local optimum numerically. 
It is interesting to note that when $T$ is small [Fig.~\ref{fig:N2_G1p5_Lx_QFI_control}(b) and Fig.~\ref{fig:N2_G1p5_T1p3_Lx}] the optimal control at small time is singular, implying taking the system close to HL-state is advantageous still. This is not the case anymore when $T$ is long when the decoherence eventually dominates.

Although the controls appear to be very different between Fig.~\ref{fig:N2_G1p5_T1p3_Lx}(b) and (c), the resulting QFI values are close. The QFI for Fig.~\ref{fig:N2_G1p5_T1p3_Lx}(b) is 2.338, only slightly smaller than 2.341 for Fig.~\ref{fig:N2_G1p5_T1p3_Lx}(c). Similar results are also observed for longer $T$. We conclude that the numerical QFI values obtained by the gradient-based algorithm are usually reliable, especially if a few different initial controls lead to close QFI values. To obtain the optimal control, the second-order conditions should be used, particularly when the majority of the control is singular.

\section{Conclusion} 
We have applied Pontryagin's Maximum Principle to analyze the parameter estimation for the dissipative quantum systems. Concretely, we consider ``twist and turn'' unitary dynamics under three different dissipation channels (depolarization, dephasing, and spin flipping), and the goal is to obtain a control protocol that maximizes QFI at a given evolution time $T$.
As a general numerical method, an efficient procedure is devised to compute switching function in dissipative systems. 
The key step is to regard $\rho$ and $\rho_\omega$ as independent dynamical variables; this augments the dynamical system but keeps the control problem in the time-invariant form so that the optimality conditions based on time-invariance hold. Using the augmented dynamics, the switching function becomes local in time and this practically crucial realization allows us to consider complicated controls.
Both the first-order condition (vanishing switching function and a time-independent control Hamiltonian) and the second-order Legendre-Clebsch condition (negative $\langle [g, [f,g]] \rangle$ and vanishing $\ddot{\Phi}$) are used to constrain the searching space of optimal controls. The second-order condition turns out to be crucial and useful for determining the values of singular control. 

For the specific ``twist and turn'' problem, we find that the optimal control is singular during the entire evolution without dissipation.
This result is verified using both first and second-order necessary conditions. 
Once the dissipation is introduced, the control can become unbounded and an amplitude constraint has to be introduced to obtain a finite solution. The onset of the instability is pinpointed using the Legendre-Clebsch condition. Three types of dissipation channels are considered. For the depolarization channel, the asymptotic QFI is zero as the system eventually goes to maximum entropy state and completely loses the coherence. For the dephasing channel, the asymptotic QFI is a non-zero value due to the decoherence-free subspace. With the optimal control, we find that QFI saturates as $T \rightarrow \infty$.  For the flipping channel the asymptotic QFI is also a non-zero value, and QFI$(T)$ has a its maximum value at a finite $T$ and then decays to a non-zero asymptotic value as $T \rightarrow \infty$. We believe this behavior is general and give detailed results for two-spin systems.  
Finally we point out two considerations that appear general beyond the specific model. First, the ability of the parameter estimation, quantified by QFI, certainly depends on the decoherence channel. To gain some intuitions about the system responses, it is instructive to consider how the HL state collapses under different decoherence channels. In the ``twist and turn'' problem, the HL state collapses to a state that is completely insensitive to the external parameter under a dephasing channel but to a state that is still sensitive to the external parameter under a flipping channel. We therefore expect and indeed find that QFI is larger under the flipping channel. 
Second, the singular control seems general in quantum problems, and the use of the second-order condition can be beneficial in determining the optimal control. 

\section*{Acknowledgment } 

We thank Y. Wang (Mitsubishi Electric Research Laboratories) for helpful discussion. 

\appendix 

\section{Time derivatives of switching function} 
Explicitly using  $\partial_\omega H = \hat{J}_z$, $\dot{\Phi}$ can be computed as 
\beq 
\begin{aligned}
\dot{\Phi} = & \text{Tr}\left\{ \hat{\lambda} \left[ [\hat{H}_0, \hat{H}_1] , \hat{\rho} \right] \right\} 
+ \text{Tr}\left\{ \hat{\lambda}_\omega \left[ [ \hat{H}_0, \hat{H}_1] , \hat{\rho}_\omega \right] \right\} + \text{Tr}\left\{ \hat{\lambda}_\omega \left[ [ \hat{J}_z, \hat{H}_1] , \hat{\rho} \right] \right\} \\
& + \underbrace{ i \left( 
 \text{Tr} \left\{ \{ \frac{d\hat{\lambda}}{dt} \} [\hat{H}_1, \hat{\rho}] \right\}
 -\text{Tr} \left\{ \lambda  \left[\hat{H}_1,  \{ \frac{d\hat{\rho} }{dt} \} \right] \right\} 
+\text{Tr} \left\{ \{ \frac{d\hat{\lambda}_\omega}{dt} \} [\hat{H}_1, \hat{\rho}_\omega] \right\}
-\text{Tr} \left\{ \lambda_\omega  \left[\hat{H}_1,  \{ \frac{d\hat{\rho}_\omega}{dt} \} \right] \right\} 
\right) }_{ \equiv E(t) }
\end{aligned}
\label{eqn:PhiDot_DM_1}
\eeq 
Note that the second line, denoted as $E(t)$, corresponds to decoherence and vanishes when $\hat{H}_1 = \hat{J}_x$ and the flipping channel; we proceed with this case and use $\hat{H}_0 = \chi \hat{J}_z^2$.  
Defining $ [\hat{J}_x, \{ \hat{J}_z, \hat{J}_y \} ] \equiv \hat{A} $,  $ [\hat{J}_z^2, \{ \hat{J}_z, \hat{J}_y \} ] \equiv \hat{B} $, and $[ \hat{J}_z, \{ \hat{J}_z, \hat{J}_y \}] \equiv \hat{E}$, one can derive   
\beq 
\begin{aligned} 
\ddot{\Phi} &= 
u(t) \left[
-\chi \text{Tr}\left\{ \hat{\lambda} \left[\hat{A} , \hat{\rho} \, \right] \right\} 
- \chi \text{Tr}\left\{ \hat{\lambda}_\omega \left[\hat{A} , \hat{\rho}_\omega \right] \right\}  
-i \text{Tr} \left( \hat{\lambda}_\omega \left[ \hat{J}_z, \hat{\rho} \right] \right) 
\right] \\
& - \left[ \chi^2 \text{Tr}\left\{ \hat{\lambda} \left[ \hat{B}, \hat{\rho} \, \right] \right\} 
+ \chi^2 \text{Tr}\left\{ \hat{\lambda}_\omega \left[\hat{B} , \hat{\rho}_\omega \right] \right\} +  \chi
\text{Tr} \left( \hat{\lambda}_\omega \left[ \hat{E} -i \{ \hat{J}_z, \hat{J}_x \}, \hat{\rho}  \right] \right)  
\right] - N_\text{damp}
\end{aligned}
\eeq  
Zero $\ddot{\Phi}$ implies the singular optimal control to be 
\beq 
\begin{aligned}
u_\text{sing} (t) &= -
\frac{ \chi^2 \text{Tr}\left\{ \hat{\lambda} \left[ \hat{B}, \hat{\rho} \, \right] \right\} 
+ \chi^2 \text{Tr}\left\{ \hat{\lambda}_\omega \left[\hat{B} , \hat{\rho} _\omega \right] \right\}  + \chi
\text{Tr} \left( \hat{\lambda}_\omega \left[ \hat{E} -i \{ \hat{J}_z, \hat{J}_x \}, \hat{\rho}  \right] \right)  + N_\text{damp}   }
{  \chi \text{Tr}\left\{ \hat{\lambda} \left[\hat{A} , \hat{\rho}  \, \right] \right\} 
+ \chi \text{Tr}\left\{ \lambda_\omega \left[\hat{A} , \hat{\rho}_\omega \right] \right\}  
+ i \text{Tr} \left( \hat{\lambda}_\omega \left[ \hat{J}_z, \hat{\rho}  \right] \right)   } \\
&= -\frac{ -D(t) + N_\text{damp} }{ -C(t) } \equiv - 
\frac{ \langle [f, [f,g]] \rangle  }{ \langle [g, [f,g]] \rangle  }
\end{aligned}
\label{eqn:u_singular_flipping}
\eeq 
With $\hat{D} = \{ \hat{J}_z, \hat{J}_y \}$, $N_\text{damp}$ is 
\beq 
\begin{aligned}
N_\text{damp} &= i \text{Tr} \left( \{ \frac{d \hat{\lambda}_\omega}{dt} \} [\hat{J}_y, \hat{\rho} ] - \lambda_\omega [\hat{J}_y, \{\frac{d \hat{\rho} }{dt} \}] \right) \\
&+ i \chi \left( 
 \text{Tr} \left\{ \{ \frac{d\hat{\lambda}}{dt} \} [\hat{D}, \hat{\rho} ] \right\}
 -\text{Tr} \left\{ \hat{\lambda}  \left[\hat{D},  \{ \frac{d\hat{\rho} }{dt} \} \right] \right\} 
+\text{Tr} \left\{ \{ \frac{d\hat{\lambda}_\omega}{dt} \} [\hat{D}, \hat{\rho}_\omega] \right\}
-\text{Tr} \left\{ \hat{\lambda}_\omega  \left[\hat{D},  \{ \frac{d\hat{\rho} _\omega}{dt} \} \right] \right\} 
\right)
\end{aligned}
\eeq 
For general channel channels, such as the dephasing, one needs to compute $\dot{E}(t)$. We just do it with bruteforce.

\section{Costate boundary condition for classical Fisher Information} 
For optimal control, the only difference between CFI and QFI is the terminal cost function which leads to a different costate boundary condition. Once the measuring basis is chosen, the classical Fisher information is given by 
\beq 
\text{CFI} = \sum_m \frac{(\partial_\omega \tilde{p}_m)^2}{ \tilde{p}_m }
= \sum_m \frac{ \tilde{\rho}_{\omega, mm}^2}{ \tilde{\rho}_{mm} } 
\equiv -\mathcal{C}_Q
\eeq 
where $\hat{ \tilde{\rho} } =  \tilde{U}^\dagger \hat{ \rho } \tilde{U} $ is the DM expressed in the measuring basis. In component form, $\tilde{\rho}_{ji} = \sum_{nm}  (\tilde{U}^\dagger)_{jm} \rho_{mn} \tilde{U}_{ni}$ where each {\em column} vector of $\tilde{U}$ is an eigenvector of $\hat{J}_x$. The costate boundary conditions are 
\beq 
\begin{aligned}
\tilde{\lambda}_{mm} &= \frac{\partial \mathcal{C}_Q}{\partial \tilde{\rho}_{mm}} = \frac{\tilde{\rho}_{\omega, mm}^2}{ \tilde{\rho}^2_{mm} } 
\Rightarrow 
\hat{ \lambda } = \tilde{U} \hat{ \tilde{\lambda} } \tilde{U}^\dagger, \\
\tilde{\lambda}_{\omega, mm} &= \frac{\partial \mathcal{C}_Q}{\partial \tilde{\rho}_{\omega, mm}} =- \frac{2 \tilde{\rho}_{\omega, mm}}{ \tilde{\rho}_{mm} } 
\Rightarrow 
\hat{ \lambda }_{\omega} = \tilde{U} \hat{ \tilde{\lambda} }_{\omega} \tilde{U}^\dagger. 
\end{aligned}
\label{eqn:costate_T_CFI_DM}
\eeq 
In the measuring basis, only diagonal components contribute to the CFI, negative of the terminal cost function.  
In the twist-and-turn problem, one introduces a phase rotation $\phi$ in the end so that the final density matrix is $e^{-i \hat{L}_z \phi} \hat{\rho} e^{+i \hat{L}_z \phi}$. In the measuring basis, 
$ \hat{ \tilde{\rho} } (\phi) =  \tilde{U}^\dagger e^{-i \hat{L}_z \phi} \hat{\rho} e^{+i \hat{L}_z \phi} \tilde{U}$  and 
$\hat{ \tilde{\rho} }_\omega (\phi) =  \tilde{U}^\dagger e^{-i \hat{L}_z \phi} \hat{\rho}_\omega e^{+i \hat{L}_z \phi} \tilde{U}$, 
based on which we get 
\beq 
\begin{aligned}
\frac{\partial \hat{ \tilde{\rho} } (\phi) }{\partial \phi} &=  \tilde{U}^\dagger e^{-i \hat{L}_z \phi} \left( -i [\hat{L}_z, \hat{\rho} ] \right) e^{+i \hat{L}_z \phi} \tilde{U} \\
\frac{\partial \hat{ \tilde{\rho} }_\omega (\phi) }{\partial \phi} &=  \tilde{U}^\dagger e^{-i \hat{L}_z \phi} \left( -i [\hat{L}_z, \hat{\rho}_\omega] \right) e^{+i \hat{L}_z \phi} \tilde{U} \\
\Rightarrow \,\, \frac{\partial \mathcal{C}_Q}{\partial \phi} &=   \sum_m \left[ \frac{ \tilde{\rho}_{\omega, mm}^2}{ \tilde{\rho}_{mm}^2 } \frac{\partial \tilde{\rho}_{mm}}{\partial \phi} - \frac{2 \tilde{\rho}_{\omega, mm}}{ \tilde{\rho}_{mm} } 
\frac{\partial \tilde{\rho}_{\omega, mm} }{\partial \phi}\right].
\end{aligned}
\eeq 
The costate boundary conditions are 
\beq 
\hat{ \lambda }= e^{+i \hat{L}_z \phi} \tilde{U} \hat{ \tilde{\lambda} } \tilde{U}^\dagger e^{-i \hat{L}_z \phi} \text{ and }
\hat{ \lambda }_\omega = e^{+i \hat{L}_z \phi} \tilde{U} \hat{ \tilde{\lambda} }_\omega \tilde{U}^\dagger e^{-i \hat{L}_z \phi}. 
\eeq  

\section{Measurement for saturated QFI$_\infty$ for $N$-spin \label{sec:CFI_Nspin}}

We consider a unitary transform $\hat{ \tilde{\rho} }(\phi) = e^{-i \hat{J}_z \phi} \hat{\rho} e^{+i \hat{J}_z \phi}$. The asymptotic $\hat{\rho}$, $\hat{\rho}_\omega$ are 
\beq 
\begin{aligned}
& \hat{\rho} = \frac{1}{2^N} \left[ (\otimes \mathbb{I})^N + (\otimes \sigma_x)^N \right], \\
& \hat{\rho}_\omega = \frac{1}{2^{N-1} \gamma} \overline{ (\otimes \sigma_x )^{N-1} \otimes \sigma_y } \\
& \text{CFI} \equiv  \sum_{m, \tilde{\rho}_{mm} >0} \frac{ \tilde{\rho}_{\omega, mm}^2 }{\tilde{\rho}_{mm}} \text{ where } \hat{ \tilde{\rho} }(\phi) = e^{-i \hat{J}_z \phi} \hat{\rho} e^{+i \hat{J}_z \phi}.
\end{aligned}
\eeq 
Keeping the non-zero terms for the measurements in the $| m_x \rangle$ basis (eigenstates of $\sigma_x$), 
\beq 
\begin{aligned}
\tilde{\rho}(\phi) &\sim \frac{1}{2^N} \left[ (\otimes \sigma_0)^N + \cos^N ( \phi ) \, (\otimes \sigma_x)^N + \cdots \right] \\
\tilde{\rho}_\omega (\phi) &\sim \frac{1}{2^{N-1} \gamma} \left[ N \cos^{N-1} (\phi) \sin(\phi) (\otimes \sigma_x)^N + \cdots
\right] 
\end{aligned}
\eeq  
The diagonal components of $\tilde{\rho}^2_\omega (\phi) $ are all $\frac{N^2 [  \cos^{N-1} (\phi) \sin(\phi) ]^2}{  (2^{N-1})^2 \gamma^2}$; those of $\tilde{\rho} (\phi) $ are $\frac{1}{2^N}(1 \pm \cos^N \phi)$. CFI is given by 
\beq 
\text{CFI}(\phi) = \frac{4}{\gamma^2} N^2 \tan^2 \phi \frac{ \cos^{2N} \phi }{ 1-\cos^{2N} \phi  }
\underset{\phi \rightarrow 0}{\rightarrow} \frac{4N}{\gamma^2}.
\eeq 
We have used $1-\cos^{2N} \phi \approx N \phi^2$ for small $\phi$. 

\bibliography{parameter_estimation}
\end{document}